\documentclass[12pt]{article}
\usepackage{amsfonts}

\newcommand{\be}{\begin{equation}}
\newcommand{\ee}{\end{equation}}
\newcommand{\bea}{\begin{eqnarray}}
\newcommand{\eea}{\end{eqnarray}}
\newcommand{\nn}{\nonumber}
\newcommand{\lb}{\left[}
\newcommand{\rb}{\right]}
\newcommand{\ac}{\mathcal{A}}
\newcommand{\bac}{\bar\mathcal{A}}
\newcommand{\ab}{\mathbb{A}}
\newcommand{\bab}{\bar\mathbb{A}}
\newcommand{\qc}{\mathcal{Q}}
\newcommand{\bqc}{\bar\mathcal{Q}}
\newcommand{\dc}{\mathcal{D}}
\newcommand{\bdc}{\bar\mathcal{D}}
\newcommand{\fc}{\mathcal{F}}

\newcommand{\hf}{{1\over 2}}
\newcommand{\ghf}{{g\over 2}}
\newcommand{\p}{\partial}
\newcommand{\iu}{I}
\newcommand{\ou}{1}

\begin{document}
\begin{titlepage}

\title{$U(2)$ instantons and leptons}

\author{M.N. Stoilov\\
{\small\it Bulgarian Academy of Sciences,}\\
{\small\it Institute of Nuclear Research and Nuclear Energy,}\\
{\small\it Blvd. Tzarigradsko Chausse\'e 72, Sofia 1784, Bulgaria}\\
{\small e-mail: mstoilov@inrne.bas.bg}}

\maketitle

\begin{abstract}
Anomalous quantisation of instantons allows 
non-trivial (anti) self-dual configurations to exist
for pure $U(2)$ gauge theory
in four-di\-men\-sional Euclidean space-time and to be used as ground state
of the model.
Six left fermions together with six right fermions and 
one complex scalar doublet describe the largest possible vacuum. 
This structure is just enough to represent the leptons and Higgs 
boson in the Standard Model.
\end{abstract}

\end{titlepage}

The aim of the present note is to construct a vacuum 
for pure $U(2)$ gauge field 
in  four-dimensional Euclidean space-time $E^4$
using instantons.
Quantisation in the vicinity of classical solution 
is a common technique in Quantum Field Theory.
When the classical solution/vacuum is not unique but
depends on some parameters the standard path integral over all fields
is replaced by a functional integral over the quantum field
(deviation from the vacuum)
 plus ordinary integrals over all parameters
on which the vacuum configuration depends \cite{t1}.
In our case the problem is even more subtle because
 we do not know the vacuum explicitly.
Instead of this we know that it satisfies some equations which are
different of the equations of motion ---
the (anti) self duality equations. 
The problem is consider in Ref.\cite{m1} for $U(1)$ gauge field
and here we deal with the $U(2)$ case with the following action
\be
{\mathbf A} =-\frac{1}{4 g'^2} \int  F_{\mu\nu}^\alpha F_{\mu\nu}^\alpha.
\label{sa}
\ee
Here $\alpha=0,1,2,3$. $F_{\mu\nu}^0$ is the field strength tensor for the
$U(1)$ potential $A_\mu^0$ 
and  $F_{\mu\nu}^a$, $a=1,2,3$ are  the field strength tensor components 
for the $SU(2)$ potential $A_\mu^a$.
\footnote{See the Appendix for details on most formulae.}

The idea is to find a class of classical solutions of the action (\ref{sa})
around which to quantise the theory.  
As the (anti) self-dual fields trivially satisfy the equations of
motion we shall use them to construct the largest possible vacuum.
The ground state thus obtained is not unique and
a proper physical parametrisation of it has to be found.
As a first step we have to rid ourselves off the gauge degrees of freedom
in the vacuum.
So, a gauge fixing of the instantons is required.
Then we rewrite the (anti-) self dual conditions and gauge fixing conditions 
in a close form which allows direct particle interpretation.
Two types of mutually commuting quaternions  are used for this purpose. 
We refer to these quaternions as $e-$ and $\xi-$quaternions respectively
with $\{e_\mu\}$ and $\{\xi^\alpha\}$ as their quaternion units.
The $e-$quaternions are connected to the space-time while
the $\xi-$quaternions are connected to the internal space.
Thus, having a gauge potential  $A_\mu^\alpha$ we  construct out of it 
four $e-$quaternion functions 
$
\ac^\alpha=e_\mu A_\mu^\alpha
$
and one bi-quaternion function
$
 \ab = \xi^\alpha \ac^\alpha.
$
We shall use also the $e-$conjugated to $\ac$ and $\ab\;$  functions
which we denote $\bac^\alpha$ and $\bab$ respectively.
All together 
\bea 
 \ab & = & \xi^\alpha e_\mu A_\mu^\alpha\nn\\
\bab & = & \xi^\alpha \bar e_\mu A_\mu^\alpha.
\eea
Here $\bar e_\mu$ is the quaternion conjugated to $ e_\mu$.
Note that there is no $\xi-$quaternion conjugation in the
definition of $\bab$.
Two conjugated first order, $e-$quaternion valued 
differential operators $\dc$ and $\bdc$
will be used as well
\bea 
\dc & = & e_\mu \p_\mu \nn\\
\bdc &= & \bar e_\mu \p_\mu.
\eea
A  $e-$quaternion function $\fc$ is called Fueter (quaternion) 
analytic if it satisfies the equation 
\be 
\dc\fc =0 \label{fa} % CHECK IT !!!!
\ee
and Fueter anti-analytic if it satisfies the equation
\be
\bdc\fc =0. \label{afa} % CHECK IT !!!!
\ee 

The Fueter analytic
and anti-analytic functions are essential for the construction of the 
 $U(1)$ nontrivial instanton vacuum.
In the $U(2)$ case the instanton equations are a nonlinear 
version of the Fueter (anti) analyticity conditions, namely,
the equation  
\be 
(\bdc + \ghf \bab)\ab = 0  \label{sdu}
\ee
describes self-dual field configuration in a certain gauge, while
the equation
\be 
(\dc + \ghf \ab)\bab = 0  \label{asdu}
\ee
describes anti self-dual field in the same gauge.
Different gauge conditions can be obtained adding an arbitrary 
$\xi-$quaternion function to eqs.(\ref{sdu},\ref{asdu}).
(Arbitrary function of $\ac^\alpha$ and $\bac^\alpha$ works fine. 
A function depending on the gauge fields derivatives can be used as well
but in this case we have to be sure that the gauge condition is
nontrivial and fixes the gauge.) 

We would like to find solutions of eqs.(\ref{sdu},\ref{asdu}) which
admit free particle interpretation.
%(i.e., with vanishing nonlinear terms).
Two such solutions are direct consequence of the $U(1)$ instanton solution
proposed in Ref.\cite{m1} and correspond to the two possible
$U(1)$ subgroup of $U(2)$ --- the $U(1)$ group in the decomposition
$U(2)= U(1)\times SU(2)$
and the Cartan subgroup of the $SU(2)$.
Each of these solutions is parametrised by a couple of
Fueter analytic and anti-analytic functions.
Additional parameters are needed in order to take
into account some features of the model.
First, we have a global $SO(3)$ invariance in the $SU(2)$ ground solution
due to the ambiguity in the choice of the Cartan subgroup.
The best way to count corresponding degrees of freedom is to introduce 
a basis in the adjoin representation.
This gives a factor of three.
Second, note that the two types of instantons can be freely combined producing 
particles with two charges.
In this case another $Z_2$ symmetry emerges. 
The symmetry is connected to the relative sign of the two  charges. 
This gives a factor of two.
Thus we end up with vacuum described by seven couples of Fueter analytic and 
anti-analytic functions.
These functions, after anomalous quantisation \cite{ds2}, correspond to seven
right and seven left Wail fermions.

%%%%% !!!!!!!!!!!!!!!!!!!!!
Let us finish with a selected list of some of the open questions
which have to be answered 
 in order to identify the vacuum just constructed with the lepton generations.
\begin{itemize}
\item 
We have excess of fermions.
A possible way out of this problem is
to quantise one of the Fueter analytic --- anti-analytic couples 
normally.
This possibility is  based on the following considerations:
Note that for quaternions the equation 
\be 
\Box \qc = 0\nn 
\ee
because of the identity
$
\dc \bdc = \Box
$
has  two solutions, namely
\bea 
\dc \qc_1 &=&0 \nn\\
 \bdc \qc_2 &=&0. \nn 
\eea 
Therefore, we can parametrise one instanton and one anti instanton 
with a complex scalar doublet.
However in this case we cannot gain stability as a consequence
of the instanton anomalous statistics.
Thus we have replaced one problem with another.
The benefit of the above construction is that now
we have a candidate for the Higgs boson.
In this case our vacuum consists of one scalar doublet and twelve Wail fermions
which coincides at least
with the type and the number of matter fields in the Standard Model.
\item 
There is no difference in our treatment between left and right chiral
fermions. 
The question why the left ones are $SU(2)$ doublets and the rights are
singlets cannot be answered.
\item 
No answer what are the particular values of the charges.
\item
Only three of the considered vacuum  solutions are stable, i.e.
we have too few stable particles.
\end{itemize}

%%%% !!!!!!!!!!!!!!!!!!

\section*{Acknowledgement}
It is a pleasure to thank Al. Ganchev for the help.

\section*{Appendix}

\vskip 10pt{\b Quaternions}

The quaternion number $\qc$ is defined as follows:
\be
\qc = e_\mu q_\mu  \label{qn}
\ee
where $q_\mu,\;\;\; \mu= 0,..,3$ are four real numbers and
$e_\mu $ are the quaternion units:
\bea
e_0 e_0 &=& 1 \nn \\
e_i e_0 &=& e_0 e_i = e_i \nn \\
e_i e_j &=& - \delta_{ij} + \epsilon_{ijk} e_k. \label{qdef}
\eea
The representation of the $e-$quaternion units we are using here is
\be
e_0  =  \ou,\;\;
e_k  =  -i\sigma_k, \;\;\; k=1,2,3 \label{e-rep}
\ee
where $\sigma_k$ are the Pauli matrices.
The $\xi-$quaternion units will be represented as real $4\times 4$ matrices 
($2\times 2$ block matrices)
\be
\xi^0 = \left(
\begin{array}{cc} 
\ou & 0\\ 0 & \ou
\end{array}\right),  
\xi^1 = \left(
\begin{array}{cc} 
0& -\iu\\ -\iu& 0
\end{array}\right),
\xi^2 = \left(
\begin{array}{cc} 
0& -\ou\\  \ou & 0
\end{array}\right),
\xi^3 = \left(
\begin{array}{cc} 
-\iu & 0\\ 0 & \iu
\end{array}\right) \label{x-rep}
\ee
where $\iu=-i\sigma_2$.
These two representations of the quaternion units together give 
well defined representation of $\mathbb{Q}^2$.

Quaternion conjugation:
\be
\overline{(\qc)}\equiv
\bqc = q_0 - e_i q_i.\label{cqn}
\ee

\vskip 10pt{\b Field strength tensor}

$U(1)$ strength tensor
\be
F_{\mu\nu}^0=\partial_\mu A_\nu^0 - \partial_\nu A_\mu^0.\label{u1fs}
\ee
$SU(2)$ strength tensor components
\be
F_{\mu\nu}^a=\partial_\mu A_\nu^a - \partial_\nu A_\mu^a +
g \lb A_\mu, A_\nu\rb^a,\;\;a=1,2,3.   \label{su2fs}   %ok
\ee

$U(1)$ charge --- $g'$.

$SU(2)$ charge --- $g'g$.

\vskip 10pt{\b Self-duality condition}

\be 
F_{\mu\nu}^\alpha=\hf \epsilon_{\mu\nu\rho\sigma}F_{\rho\sigma}^\alpha,
\;\; \alpha=0,1,2,3       %ok
\ee
($\epsilon$ is the totally anti symmetric tensor.)
For each $\alpha$ these are three independent conditions.
In the $U(1)$ case they are
\be 
\partial_0 A^0_i- \partial_i A^0_0 -
 \epsilon_{ijk}\partial_j A^0_k =0    %ok
\ee 
and for $SU(2)$ they are
\be 
\partial_0 A^a_i- \partial_i A^a_0 -
 \epsilon_{ijk}\partial_j A^a_k +
g \epsilon^{abc}\left(A_0^b A_i^c -
\hf \epsilon_{ijk} A_j^b A_k^c\right)=0.    %ok
\ee 
Here we have used
$
\lb A_\mu, A_\nu\rb^a = \epsilon^{abc}A_\mu^b A_\nu^c
$.

\vskip 10pt{\b Some algebra}

Eq.(\ref{sdu}) is equivalent to the following  equations
\bea 
0&=& \p_\mu A_\mu^0 + \ghf \left((A^0)^2-(A^a)^2\right)\nn\\
0&=& \p_\mu A_\mu^a + g A^0\cdot A^a \nn\\
0&=& \p_0 A_i^0 - \p_i A_0^0 - \epsilon_{ijk}\p_j A_k^0\nn\\
0&=& \p_0 A_i^a - \p_i A_0^a - \epsilon_{ijk}\p_j A_k^a
+g\epsilon^{abc}(A_0^b A_i^c - \hf \epsilon_{ijk} A_j^b A_k^c)
\label{a-sd-exp}
\eea
and so, it describes self dual gauge field (last two of eqs.(\ref{a-sd-exp}))
in a gauge which is a variant of the Lorentz gauge (first two equations).
Analogously, eq.(\ref{asdu})
is equivalent to the following equations
\bea 
0&=& \p_\mu A_\mu^0 + \ghf \left((A^0)^2-(A^a)^2\right)\nn\\
0&=& \p_\mu A_\mu^a + g A^0\cdot A^a \nn\\
0&=& \p_0 A_i^0 - \p_i A_0^0 + \epsilon_{ijk}\p_j A_k^0\nn\\
0&=& \p_0 A_i^a - \p_i A_0^a + \epsilon_{ijk}\p_j A_k^a
+g \epsilon^{abc}(A_0^b A_i^c + \hf \epsilon_{ijk} A_j^b A_k^c)
\label{a-asd-exp}
\eea
and it describes anti self dual field in the same gauge 
as in eq.(\ref{a-sd-exp}).
The different lines in eqs.(\ref{a-sd-exp},\ref{a-asd-exp}) correspond
to the coefficients (up to sign) for the quaternion units
$1$, $\xi^a$, $e_i$ and $\xi^a e_i$, so
in both cases the gauge conditions
are pure $\xi-$quaternions while the (anti) self-dual conditions
are imaginary $e-$quaternions.

\vskip 10pt{\b Simple instanton solution I}

Let $\ac^a=0\;\;\forall a$ and $\ac^0\neq 0$. 
In this case eq.(\ref{sdu}) reduces (after suitable change of the gauge)
to Fueter anti analyticity condition (eq.(\ref{afa})) for $\ac^0$
while eq.(\ref{asdu}) goes to  Fueter analyticity condition (eq.(\ref{fa})).
The anomalous quantised solutions of equations (\ref{fa},\ref{afa}) 
 are massless fermions \cite{ds2}.

\vskip 10pt{\b Simple instanton solution II}

Pick up an index $a$, say $a=\mathbf{a}$. 
Let $\ac^\alpha=0\;\;\forall \alpha\neq \mathbf{a}$ and
$\ac^\mathbf{a}\neq 0$.
Again  eq.(\ref{sdu}) reduces to eq.(\ref{afa}) for $\ac^\mathbf{a}$ and
eq.(\ref{asdu}) reduces to eq.(\ref{fa}).

There is a global $SO(3)$ covariance for solution II:
If $\ac^a$ is a solution, so is the field 
$
\ac '^a = U^{ab}\ac^b
$ 
where $U^{ab}$ is $SO(3)$ matrix.

\end {document}